\documentclass[12pt]{article}
\usepackage{xspace,amssymb,amsmath,helvet,graphicx,url,setspace}

\begin{document}
\newcommand{\dee}{\,\mbox{d}}
\newcommand{\naive}{na\"{\i}ve }
\newcommand{\eg}{e.g.\xspace}
\newcommand{\ie}{i.e.\xspace}
\newcommand{\pdf}{pdf.\xspace}
\newcommand{\etc}{etc.\@\xspace}
\newcommand{\PhD}{Ph.D.\xspace}
\newcommand{\MSc}{M.Sc.\xspace}
\newcommand{\BA}{B.A.\xspace}
\newcommand{\MA}{M.A.\xspace}
\newcommand{\role}{r\^{o}le}
\newcommand{\signoff}{\hspace*{\fill} Rose Baker \today}
\newenvironment{entry}[1]%
{\begin{list}{}{\renewcommand{\makelabel}[1]{\textsf{##1:}\hfil}%
\settowidth{\labelwidth}{\textsf{#1:}}%
\setlength{\leftmargin}{\labelwidth}
\addtolength{\leftmargin}{\labelsep}
\setlength{\itemindent}{0pt}
}}%
{\end{list}}
\title{Outliers in meta-analysis: an asymmetric trimmed-mean approach}
\author{Rose Baker\\School of Business\\University of Salford, UK}
\maketitle
\begin{abstract}
The adaptive asymmetric trimmed mean is a known way of estimating central location, usually in conjunction with the bootstrap.
It is here modified and applied to meta-analysis, as a way of dealing with outlying results by down-weighting
the corresponding studies. This requires a modified bootstrap and a method of down-weighting studies, as opposed to removing single observations.
This methodology is shown in analysis of some well-travelled datasets to down-weight outliers in agreement with other methods, and Monte-Carlo studies show
that it does does not appreciably down-weight studies when outliers are absent. Conceptually simple, it does not make parametric assumptions about the outliers.

\end{abstract}
\section*{Keywords}
meta-analysis, random effect, outlier, trimmed mean, bootstrap.

\section{Introduction}
In  meta-analyses, the results of studies often disagree with each other by more than their quoted standard errors would indicate.
A common approach in this situation is to use a random-effects model, where  each study effectively measures a different value of the effect to be estimated.
The random effect is assumed to be normally distributed, and in medicine covers such causes of inter-study variation as differences in patient mix and in operational procedure.
However, even after adding a random effect, results can still be abnormally large or small, and it is not clear what should be done with such studies.
This is the problem of outliers, one of several that plague meta-analysis, along with publication bias and temporal trends in treatment effect.

Statisticians do not like removing outliers simply because they look wrong, and there have been several statistical attempts to address the outlier problem more systematically.
Baker and Jackson (2008) fitted symmetric long-tailed distributions, and {\em ibid} (2016) proposed some skew long-tailed models
of the random effect. Similarly, Beath (2014) used a mixture of normal distributions to model the random effect. The reasoning in these articles is that there is no compelling reason why the random effect should be normally-distributed, unlike the statistical error,
whch should be normally-distributed by virtue of the central limit theorem. 
Fitting long-tailed distributions has the effect of down-weighting outliers,
but the drawback is that modelling is required and the fitted distribution requires extra parameters to be flexible enough. 

Viechtbauer and Cheung (2010) presented a methodology for identifying outliers, 
which is equivalent to testing for a shift in mean.
Similarly, Gumedze and Jackson (2011) proposed a REML (restricted maximum likelihood estimation) based method whereby the random effect variance can be inflated for a study.
A likelihood ratio test then enables outliers to be identified, and such outliers are 
reduced in effect. This approach again has pros and cons, drawbacks being that some judgement is still required to
identify possible outliers, and that the method becomes cumbersome with more than one outlier. 

Julious and Whitehead (2011) gave graphical methods for identifying outliers,
and Mavridis {\em et al} (2017) gave a stepwise procedure  in which studies are progressively added. Lin {\em et al} (2017) proposed new measures of heterogeneity based on medians
that are robust against outliers.
Despite all this research, there is as yet no fully `automatic' way of dealing with outliers, that does not make modelling assumptions about them or require judgement, and this work tries to fill that gap.

The approach here is heuristic rather than theoretical; the method of dealing with outliers described is largely nonparametric, and is motivated by the use of the asymmetric trimmed mean
outside meta-analysis.
The trimmed mean has proportions of observations $\alpha_L, \alpha_U$ trimmed from the lower and upper ends of the curve. 
Trimmed means have been used to obtain robust estimates of central location in the presence of outlying observations, \eg Aleem and Lahiani (2011), Sharma and Bicchel (2015).
The optimum proportions of the observations to be trimmed can be chosen adaptively, to minimise the variance of the trimmed mean. 
Such a trimmed estimator is appropriate because it  is the most reproducible estimate of the treatment effect that can be derived from the data.
The bootstrap offers a convenient way to 
compute the variance of the trimmed mean and so to choose the optimal proportions (\eg L\'{e}ger and Romano, 1990, Efron and Tibshirani, chapter 18, 1993).
Clearly, an asymmetric trimmed mean rather than a symmetric one is needed, when outliers can give high or low results, or both.

In meta-analysis, we do not have single observations, but studies, which necessitates a development of the trimming method and also of the bootstrap methodology.

The next sections discuss the trimming model and the bootstrap methodology used.
Finally, the method is applied to some datasets that have been used before to study outliers, and a simulation study is done.
\section{The trimming model}
Let the estimated effect from the $i$th of $n$ studies be $y_i$ and the unknown mean $\mu_i$, with the variance $\sigma_i^2$ assumed known as is customary in meta-analysis (but in reality
of course it is an estimate). The trimmed mean described earlier must be adapted to meta-analysis, where we deal with studies rather than single observations, and
the solution adopted here is to have lower and upper bounds $B_L, B_U$ within which the true mean of a study should lie. For each study, given $y_i, \sigma_i$, the probability $P_i$ that the mean lies in $(B_L, B_U)$ can be computed.
The notional number of observations in study $i$ is then multiplied by $P_i$ to give the expected number of admissible observations, \ie the variance $\sigma_i^2 \rightarrow \sigma_i^2/P_i$.
Thus studies where the mean could be in the tail have variance inflated and are down-weighted.

The procedure is, for a range of values of $\alpha_L, \alpha_U$, and for a  bootstrapped sample, to compute from $\alpha_L, \alpha_U$ the 
corresponding $B_L, B_U$, which will be different for each bootstrapped sample, and then to down-weight the studies accordingly,
and so compute the mean effect estimate $\hat\theta$. This is done for a large number $N$ of bootstrapped samples and the variance of $\hat\theta$ computed.
Finally, the values $\alpha_L, \alpha_U$ for which the variance of $\hat\theta$ over the bootstrap samples is minimum are used to estimate $\theta$ from the observed data.

Suppose that only studies whose means $\mu_i$ lie in the range $(B_L, B_U)$ are to be included.
Then $y_i=\mu_i+\sigma_i\epsilon$ where $\epsilon \sim N[0,1]$.
The probability that $\mu_i > B_U$ is 
\[\text{Prob}(\mu_i > B_U)=\text{Prob}(y_i-\sigma_i\epsilon > B_U)=1-\Phi(-(y_i-B_U)/\sigma_i)=\Phi((y_i-B_U)/\sigma_i),\]
where $\Phi$ is the standard normal distribution function.
Similarly, the probability that $\mu_i < B_L$ is $1-\Phi((y_i-B_L)/\sigma_i)=\Phi((B_L-y_i)/\sigma_i)$.
The probability $P_i$ that $B_L \le \mu_i \le B_U$, so that the study should be included, is 
\[P_i=\Phi((B_U-y_i)/\sigma_i)-\Phi((B_L-y_i)/\sigma_i).\]

We work in terms of $\alpha_L, \alpha_U$,
and compute  $B_L, B_U$ from $\alpha_L, \alpha_U$ by solving $\sum_{i=1}^n v_i\Phi((y_i-B_U)/\sigma_i)=\alpha_U$,
$\sum_{i=1}^n v_i\Phi((B_L-y_i)/\sigma_i)=\alpha_L$, where $v_i=\frac{1/\sigma_i^2}{\sum_{j=1}^n1/\sigma_j^2}$.
This is done by a robustified Newton-Raphson method, where the binary chops method was used to obtain a starting value for the Newton-Raphson iteration; any robust equation-solver could be used instead.

The underlying meta-analysis methodology can be either fixed or random effects, but one would expect to use a random-effects model where outliers are suspected. 
In a random-effects model, the between-study variance $\tau^2$ of $\theta$ is added to the statistical variance $\sigma_i^2$ of a study.
Merely adding a normally-distributed error is not sufficient where there may be outliers, but here, a normally-distributed random effect is used,
and extreme results are shaved away.
\section{The bootstrap}
To bootstrap under the random-effects model, the `error bootstrap' from Van Den Noortgate and Onghena (2005) was used, with two adjustments, which will be described.
It is crucial to use a nonparametric bootstrap, in which one attempts to reconstruct the means $\mu_i$ and then add a random error, rather than a parametric bootstrap,
in which fresh $\mu_i$ are randomly generated from the fitted estimates $\hat\theta$ and $\hat{\tau}^2$. The nonparametric bootstrap then yields bootstrapped datasets
that preserve the outliers, and enable the $\alpha_L, \alpha_U$ to be found that are optimal for the original dataset. From the parametric bootstrap, one would expect simply to find that $\alpha_L=\alpha_U=0$,
which is what was observed.

With the estimated random effect variance ${\hat\tau}^2$ computed, \eg using the DerSimonian and Laird (DSL) method (ibid, 1986, or \eg Hartung {\em et al}, 2008), empirical Bayes estimates of the study means $\mu_i$ are computed.
There are two estimates of the study mean $\mu_i$: the observation $y_i$, and $\hat{\theta}$, which has variance estimated as $\hat{\tau}^2$. Combining these gives the Empirical Bayes estimate
\[\tilde{\mu}_i=\frac{\hat{\tau}^2 (y_i-\hat{\theta})}{\hat{\tau}^2+\sigma_i^2}+\hat{\theta}.\]
The residuals are then computed as
\begin{equation}\tilde{\eta}_i=\frac{\hat{\tau}^2 (y_i-\hat{\theta})}{\hat{\tau}^2+\sigma_i^2}.\label{eq:eta}\end{equation}
Let the usual sample variance of these $n$ shrunken residuals be $V(\hat{\tau})$.
Because these are shrunk towards zero, $V(\hat{\tau}) < \hat{\tau}^2$, and they are therefore `reflated' by multiplying by $\hat{\tau}/V(\tau)^{1/2}$
to have the correct estimated variance $\hat{\tau}^2$.
Once $\hat\theta$ is added, these are the  best estimates of the means $\mu_i$,
and $n$ of them are selected with replacement.
Each mean then has a random error $\sigma_i\epsilon$ added, where $\epsilon \sim N[0,1]$. This is a small adjustment to the method given
in Van Den Noortgate and Onghena (2005), in that here normally-distributed random errors are added to the means rather than bootstrap residuals.
A bootstrapped value $x_i^\prime$ is thus given by
\[x_i^\prime=(\hat{\tau}/V(\hat{\tau})^{1/2})\tilde{\eta}_j+\hat{\theta}+\epsilon\sigma_j,\]
where $\epsilon \sim N[0,1]$ and the $j$th study is randomly chosen as the $i$th.

A bigger adjustment to the method of Van Den Noortgate and Onghena is changing the method of reflation. Without reflation, the calculated means for the studies have a variance $< \hat{\tau}^2$, but with it, large studies are overinflated.
Thus a study result $y_i$ from a large study with a tiny variance $\sigma_i^2$ is hardly shrunk at all by (\ref{eq:eta}) to give the residual ${\tilde\eta}_i$, and is then enlarged by an inflation factor of typically 2 or 3.
Hence large studies take more extreme values in the bootstrapped datasets than in the original. 
A solution to this problem is to shrink the residuals $y_i-\hat{\theta}$ by a smaller amount,
using a larger variance $\phi^2$ such that 
\[\tilde{\eta}_i(\phi)=\frac{\phi^2 (y_i-\hat{\theta})}{\phi^2+\sigma_i^2},\]
where $V(\phi)=\hat{\tau}^2$. Now there is no further reflation, and large studies are not over-emphasized in the bootstrapped sample.
A bootstrapped value is given by
\[x_i^\prime=\tilde{\eta}(\phi)_j+\hat{\theta}+\epsilon\sigma_j.\]

The variance $\phi^2$  can be computed from solving  $V(\phi)-\hat{\tau}^2=0$. A robust method to find $\phi^2$ is to
find lower and upper bounds, then use (say) 6 steps of a binary chop algorithm, and finally use the Newton-Raphson method. Alternatively, a single-variable equation-solver can be used.

\section{Data analysis}
Eight datasets that have already been studied for outliers were used, and brief details are given in table \ref{table0}.
\begin{table}[h]
\begin{tabular}{llll} \hline
Name & Description & Sign & Studies \\ \hline
Paroxetine &Paroxetine vs placebo for depression& +ve & 23 \\ \hline
Mg & Magnesium for acute myocardial infarction & ~-ve & 15 \\ \hline
CDP & CDP-choline for cognitive disturbances & +ve & 10 \\ \hline
Fluoride& Toothpaste for dental caries& -~ve & 70 \\ \hline
Pravastatin&Lowering cholesterol&+ve &64 \\ \hline
Aspirin&Oral aspirin for acute postoperative pain& +ve & 63 \\ \hline
Academic&Writing-to-learn intervention & +ve & 26 \\ \hline
Jobperf &Correlation between commitment and job performance & +ve & 61 \\ \hline
\end{tabular}
\caption{\label{table0}Details of the 8 datasets used. The sign is the beneficial direction.}
\end{table}

The fluoride toothpaste, CDP-choline, pravastatin and paroxetin datasets are all described in Baker and Jackson (2016).
The aspirin dataset is described in Baker and Jackson (2008), while the effect of writing-to-learn interventions on academic
achievement dataset and the job-peformance dataset were studied by Viechtbauer and Cheung (2010) and the academic dataset also by Mavridis {\em et al} (2017). 
Finally, the effect of magnesium on myocardial infarction dataset has been discussed by  Higgins and Spiegelhalter (2002).

Computations were done with a purpose-written fortran program, available from the author. This  will run under a freely-available fortran compiler and some further information is given in the appendix.
The number $N$ of bootstrap replications was taken as 10000: going to 100000 changed the results only very slightly.
The strategy was to compute the bootstrapped variance for a large number of values of $\alpha_L, \alpha_U$, from a single set of $N$ bootstrapped datasets. 
\section{Results}
Table \ref{table1} summarizes the results of fitting the DSL random effects model to the 8 datasets,
together with the results of the trimmed method. In general, one sees that trimmed datasets have as expected lower values of the random effect variance $\tau^2$,
and that `bagged' (bootstrap-averaged) means and standard errors agree fairly well with the parametric values. 
\begin{table}[h]
\begin{tabular}{lllllll} \hline
Dataset & $\alpha_L$ & $\alpha_U$ & $\hat{\theta}$& $\tilde{\theta}$& $\tau$ & $I^2$  \\ \hline
Paroxetine & 0&0& 3.356 (.393)&3.321(.408)&1.671 & 87.1\% \\ \hline
&0&0.425&1.945 (.173)&2.023 (.290) &0&0\% \\ \hline \hline
Mg & 0 & 0 & -0.605 (.181)  & -.477 (.280)& .416 & 64.4\% \\ \hline
&0.780&0&0.054 (.051)&-0.003 (.168)&0&0\% \\ \hline \hline
CDP & 0 & 0 & 0.378 (.138)&.320  (.127)& .339 & 67.5\% \\ \hline
&0&0.340&0.149 (.076)&0.146 (.089)&0&0\% \\ \hline \hline
Fluoride & 0 & 0 & -0.303 (.020)  & -.287 (.017)& .133 &  70.1\% \\ \hline
& 0.031 & 0 & -0.269 (.015) & -.274 (.017) & .075 & 42.04\% \\ \hline \hline
Pravastatin &  0 & 0 & 29.50 (.418)  & 29.48 (.415)& 3.34 & 99.98\% \\ \hline
& 0.99& 0 & 35.43 (.416) & 34.11 (.286)& 1.23 & 88.39\% \\ \hline \hline
Aspirin & 0 & 0 & 1.257 (.082)  & 1.212 (.084)& .200 & 9.84\% \\ \hline
&0&0.004&1.233 (.079)& 1.203(.084) & .127 & 4.18\% \\ \hline \hline
Academic & 0 & 0 & 0.251 (.070) & 0.232 (.071) & .238 & 49.7\% \\ \hline 
& 0 & 0.077& 0.161 (.054) & 0.172 (.070) & .096 & 12.91\% \\ \hline\hline
Jobperf&0&0&0.188 (.020) & 0.185 (.019) & .131 & 79.43\% \\ \hline
&0&0.0125&0.178 (.019) & 0.179 (.019) & .121 & 76.53\% \\ \hline
\end{tabular}
\caption{\label{table1}Results for analysis of the 8 datasets used, showing the optimum trimming proportions, the parameter estimate with standard error, 
the bootstrap (bagged) mean with bootstrap standard error, the random effect $\tau$, and the Higgins and Thompson $I^2$.
The upper line for each dataset has $\alpha_L, \alpha_U$ set to zero.}
\end{table}
Results for the paroxetine analysis agree well with Baker and Jackson (2016), who obtained $\hat{\theta}=2.223 \pm 0.223$. As in their work, the estimated effect
is now lower than the fixed-effects estimate. The same studies have been heavily down-weighted. There is a drastic reduction of the estimated effect.

The magnesium dataset is unusual (see \eg Higgins and Spiegelhalter, 2002) and is included simply to satisfy curiosity as to how the trimming method would deal with it. The very large ISIS-4 trial shows no treatment effect,
but several small studies do show one; there is now believed to be little or no treatment effect.
Fixed and random effect models give very different answers, and the random effect model gives the `wrong' answer.
Here, the smaller bagged mean and larger standard error already show a reduction in the effect ($p=0.044$), but trimming removes much of the small
studies, and  gives a result similar to the (correct) fixed-effects model, $\hat{\theta}=0.0166 \pm 0.035$. This result may be fortuitous, but is perhaps encouraging.

The treatment effect $\hat\theta$ changed drastically for the CDP dataset, where there is one obvious outlier, the third study.
Table \ref{table2} shows that the weight of this study was reduced to zero, while the other studies retained much of their weight.
Study 4 was reduced to 20\% of its weight. If the only outlier is study 3, then this is `collateral damage'. One cannot remove study 3 without
reducing the weights of other studies that find large effects.  The estimated effect agrees with Baker and Jackson (2016).
\begin{table}[h]
\begin{tabular}{cccc} \hline
obs. &  effect &   sd  &	proportion kept \\ \hline
   
     1  &    0.2600 &     0.3827    &  0.5546 \\ \hline
    2   &   0.0100&      0.2092      &1.0000\\ \hline
    3    &  2.2200 &     0.4107      &0.0000\\ \hline
    4     & 0.5800  &    0.3699      &0.2043\\ \hline
    5      &0.3400   &   0.3699     & 0.4520\\ \hline
    6&      0.3300    &  0.2857    &  0.4305\\ \hline
    7 &     0.1400    &  0.2398   &   0.7614\\ \hline
    8  &    0.1500     & 0.0969  &    0.9427\\ \hline
    9  &    0.5000      &0.2704 &     0.1904\\ \hline
   10   &   0.1300      &0.2117&      0.8084\\ \hline
\end{tabular}
\caption{\label{table2}Effect sizes, standard errors and proportions retained for the CDP dataset.}
\end{table} 
The fluoride toothpaste study has several outliers. Here many studies were reduced in weight, and the effect was 
to decrease the (negative) treatment effect by just over 11\%. The effect size agrees well with a value of -0.273 found by Baker and Jackson (2016).

The pravastatin dataset shows a decline of effect size with time,
which was neglected in this work. Here nearly all the data has been trimmed away, with only the early (large) effects retained. The results certainly flag up a dataset that is problematical
and requires further attention.

In the aspirin dataset, five high-effect studies were down-weighted to about 70\%. Baker and Jackson (2008) found no effect on $\hat\theta$ of fitting long-tailed distributions, because they only considered
symmetric distributions. The reduction in treatment benefit is only about 2\%.

The effect size decreased drastically for the academic dataset, where the weight of study 25 was reduced to 1\% of its original value.
This agrees with the findings of Viechtbauer and Cheung (2010) and Mavridis {\em et al} (2017).

Finally, the results for the job-performance dataset show only a small change in effect on down-weighting outliers, as found by Viechtbauer and Cheung (2010).
As in their analysis, study 56 has been down-weighted (to 10\% here), but study 8, which they also removed, was hardly down-weighted.
Overall, this methodology agrees well with previous work on outliers.

To study the effect of using this methodology when there are no outliers, random datasets were generated from each of the 8 datasets, retaining the fitted random-effects variance $\tau^2$
and the study variances $\sigma_i^2$, but generating fresh study results $y_i$ from the normal random-effects model, \ie generating resamples using the parametric bootstrap. 
Without loss of generality, the treatment effect was set to zero.
Results are shown in table \ref{table3}. In 4 cases, $\alpha_L=\alpha_U=0$, so the method does not change the original results at all. In other cases, a small amount of trimming was found optimal,
but again the fitted treatment effect hardly altered, and was always below its standard error in magnitude, so the method never produced a treatment effect when there was none.
The CDP-choline dataset has the smallest sample size, and here $\alpha_L, \alpha_U$ were nonzero for the Monte-Carlo dataset.
Ten such random datasets were analysed, and the mean value of $\alpha_L, \alpha_U$ was $\alpha_m=0.052$. There is low-high symmetry, so the average $\alpha_L$ was averaged with the average $\alpha_U$
to obtain this statistic. This could be taken as a measure of bias in the computation of the $\alpha_L, \alpha_U$, in that even random datasets can appear to have outliers
when the number of studies is small. This suggests a correction for small numbers of studies, that $\alpha_L \rightarrow \max(\alpha_L-\alpha_m, 0)$, and similarly for $\alpha_U$.
\begin{table}[h]
\begin{tabular}{lllllll} \hline
Dataset & $\alpha_L$ & $\alpha_U$ & $\hat{\theta}$& $\tilde{\theta}$& $\tau$ & $I^2$  \\ \hline
MC-Paroxetine & 0&0& .1022 (.3173)&.1369 (.3168)&1.27&79.66\% \\ \hline
&0& 0.0013 & .0976 (.3176)&.1322 (.3167)& 1.27&79.63\% \\ \hline \hline
MC-Mg & 0 & 0 & -.1043 (.1497) & -.1299 (.1875)&.2916&47.06\% \\ \hline
 & 0 & 0 & -.1043 (.1497) & -.1299 (.1875)&.2916&47.06\% \\ \hline
MC-CDP & 0 & 0 &.0532 (.1047)&.0554 (.0953)&.2036&42.88\% \\ \hline
&.056&.08&.0569 (.0770)&.0510 (.0913)&.0467&2.96\% \\ \hline \hline
MC-Fluoride & 0 & 0 & -.0005 (.0184)&-.0032 (.019)&.115&63.70\% \\ \hline
 & 0 & 0 & -.0005 (.0184)&-.0032 (.019)&.115&63.70\% \\ \hline
MC-Pravastatin &  0 & 0 & .1518 (.2474)&.1269 (.2485)&1.9733&99.95\% \\ \hline
&  0 & 0 & .1518 (.2474)&.1269 (.2485)&1.9733&99.95\% \\ \hline
MC-Aspirin & 0 & 0 & -.0005 (.0184)&-.0032 (.0190)&.115&63.70\% \\ \hline
& 0 & 0 & -.0005 (.0184)&-.0032 (.0190)&.115&63.70\% \\ \hline
MC-Academic & 0 & 0 & -.011 (.0464)&-.0076 (.0552) & .1144&18.58\% \\ \hline 
& 0 & 0.0013& .002 (.0533)&-.0086 (.0552) & .1115&17.86\% \\ \hline\hline
MC-Jobperf&0&0&-.011 (.0464)&-.0076 (.0552)&.1144&18.58\% \\ \hline
&0&0.0013&.011 (.017)&.0139 (.017)&.1084&72.65\% \\ \hline
\end{tabular}
\caption{\label{table3}Results for analysis of the Monte-Carlo versions of the 8 datasets used, showing the optimum trimming proportions, the parameter estimate with standard error, 
the bootstrap (bagged) mean with bootstrap standard error, the random effect $\tau$, and the Higgins and Thompson $I^2$.}
\end{table}

\section{Conclusions}
This work has shown that the concept of an asymmetric trimmed mean can be usefully applied in meta-analysis, with the aid of the bootstrap.
Results for the 8 datasets agree closely with those obtained using parametric models (\eg Baker and Jackson, 2016), regression diagnostics (Viechtbauer and Cheung, 2010)
or stepwise procedures (Mavridis {\em et al}, 2017).
This method however does not have the arbitrariness of model choice of parametric modelling of long-tailed distributions.
The Dersimonian and Laird method of meta-analysis has been used, but any method could be plugged into this methodology.
For simplicity, meta-regression on covariates has been ignored in this work, but again can be easily included. The bootstrap is then modified so that the
bootstrapped studies retain their covariates. 
It is also possible to relax the usual assumption that the $\sigma_i$ are known exactly. For example, a t-distribution could be assumed.

The statistic $\alpha_L+\alpha_U$ can be used as an index of outlier prevalence; here the value is very high for several datasets.
After seeing that the dataset needs trimming, one could simply accept the trimmed result as giving a more precise effect estimate.
It would also however be possible to entirely remove studies that have been drastically down-weighted, or to treat them as outliers using the method of Gumedze and Jackson (2011).
The proportion of a study retained is the probability that its mean $\mu_i$ is `admissible' and so could be taken as a probability that the study is not an outlier.

How then should this methodology be best used? It is probably best to carry out a conventional analysis, and then to apply the asymmetric trimming method
as a sensitivity analysis if there is any suspicion that one or more studies might be outliers. If the results differ much, and if $\alpha_L+\alpha_U$
is large, the results of the asymmetric analysis should be used. Monte-Carlo studies show that the method will not produce a treatment effect when there is none.

The statistical error on $\alpha_L, \alpha_U$ is not included in the computed standard error of the treatment effect $\hat\theta$. To get a 95\% confidence interval for $\theta$,
one would need to use a double bootstrap. In this, $\alpha_L, \alpha_U$ would be estimated for each bootstrapped sample using a second level of bootstrap,
and the percentiles of $\hat\theta$ taken from the first level of bootstrapped datasets where each sample has had the optimal values of $\alpha_L, \alpha_U$ applied.

The approach here has been largely empirical, and the properties of the asymmetric trimmed estimator as used here could be studied more theoretically, and also using more datasets.
The error bootstrap ( Van Den Noortgate and Onghena, 2005) has been slightly improved in this article, so that the calculated mean effect for very large studies
is now very close to the observed effect. Further work on this type of bootstrap would be useful.

If practitioners find this method helpful, it could  be ported from the fortran prototype to the R language as a package.
\section*{Highlights}
\begin{itemize}
\item Outlying studies can greatly distort the results of meta-analysis;
\item Several approaches have been tried to deal with this problem, \eg fitting parametric models;
\item A largely nonparametric method of mean trimming was adapted to work in meta-analysis;
\item The adaptation was not automatic but required some new mathematics;
\item Results show good performance that agrees with previous work, without making a lot of modelling assumptions.
\end{itemize}

\clearpage
\section*{Appendix: code}  
Currently, the code is a fortran95 program that reads a `control' file of commands, and the datafile.
It uses the NAG library for random-number generation and for computing the normal distribution function.
Results are written to a listing file.
It may be obtained by emailing the author, together with a sample dataset and a listing file.

It would be easy to render the code independent of the NAG library, by replacing random-number routines 
G05KFF, G05SAF, and the normal distribution function G01EAF.

Conversion to R would be more work, but fortran code can be called from R, so this too is feasible.
\end{document}